\documentclass[english]{article}
\usepackage{graphicx,color}
\def\no{\noindent}
\def\bc{\begin{center}}
\def\ec{\end{center}}

\def\beq{\begin{equation}}
\def\eeq{\end{equation}}

\def\br{{\bf r}}
\def\bq{{\bf q}}
\def\bk{{\bf k}}

\begin{document}

%Title of paper
\title{Anderson localization in a two-dimensional random gap model\\
%in two-dimensional random systems with particle-hole symmetry\\
%{\small (al13.tex)}
}

\author{A. Hill and K. Ziegler\\
%\affiliation{Institut f\"ur Physik, Universit\"at Augsburg\\
Institut f\"ur Physik, Universit\"at Augsburg\\
D-86135 Augsburg, Germany}
%\date{\today}

\maketitle

%\begin{abstract}
\no
Abstract:

\no
We study the properties of the spinor wavefunction in a strongly disordered environment
on a two-dimensional lattice. By employing a transfer-matrix calculation we find that
there is a transition from delocalized to localized states at a critical value of the disorder
strength. We prove that there exists an Anderson localized phase with exponentially decaying
correlations for sufficiently strong scattering.
Our results indicate that suppressed backscattering is not sufficient to prevent Anderson localization 
of surface states in topological insulators.
\vskip0.5cm

\no

\section{Introduction}

The classical approach to randomly scattered particles leads to diffusion, where random scattering originates 
either from particle-particle collisions (e.g., in a gas) or from collisions with (static) impurity scatterers. 
In quantum systems, however, diffusion appears only for weak disorder whereas it is destroyed due to Anderson 
localization at stronger randomness \cite{anderson58,abrahams79}. This effect is particularly strong in 
low-dimensional systems, such as two-dimensional graphene sheets or the surface of topological insulators.
The scaling approach to generic random scattering \cite{abrahams79}
states that diffusion is entirely suppressed by Anderson localization for dimension $d\le 2$.
On the other hand, it has been argued that Anderson localization is prevented on the surface of topological 
insulators due to suppressed backscattering \cite{zhang11,culcer11}.

Inspired by the recent observation of metallic behavior (i.e. diffusive or even ballistic transport) in
disordered two-dimensional systems (graphene) \cite{novoselov05,zhang05}, a general discussion of diffusion and
localization of quantum particle is required, which takes into account a spinor structure of the wavefunction.
Two possibilities have been considered, namely ballistic transport for finite systems \cite{rosenstein10,morandi11} 
and diffusive transport for infinite systems \cite{ziegler97}. Diffusion is related to long ranged correlations,
which is usually caused by spontaneous symmetry breaking \cite{ziegler97,wegner80}. This behavior might
be restricted to the regime of weak scattering, since strong scattering is capable to localize particles. The aspect 
of weak localization is ignored here on purpose because it has its own problems \cite{andoetal,khveshchenko06}. 
This will be discussed in a separate
paper. Instead, we will focus in the following mostly on the case of strong scattering. This is motivated by recent numerical
studies, which have indicated that there is a transition to a localized phase at sufficiently strong
disorder \cite{hill12,yamakage2012}. Here we will analyze details of the transition in terms of the scaling behavior
of the localization length for strips of finite width. Moreover, the infinite system will be treated analytically  within a 
strong scattering expansion. The latter provides a rigorous proof for exponential localization, supporting the
numerical results at strong disorder. We study a random gap model with linear spectrum 
(2D Dirac fermions), but our methods can be easily applied to other systems as well.

\section{Model}

We consider the surface Hamiltonian of a topological insulator with bulk inversion symmetry of momentum $\bk$ 
\cite{zhang11,yamakage2012,bernevig,yamakage2011}
\beq
H=\pmatrix{
h(\bk) & 0 \cr
0 & h^*(-\bk) \cr
}, \ \ \ h(\bk)=\hbar\pmatrix{
C+M-(D+\delta)k^2 & v_F(k_x+ik_y) \cr
v_F(k_x-ik_y) & C-M-(D-\delta)k^2 \cr
}
\label{eq:plain-hamiltonian}
\eeq
This Hamiltonian consists of a pair of massive Dirac Hamiltonians $h(\bk)$, $h^*(-\bk)$. It should be noticed that this 
Hamiltonian reads in coordinate space
\[
H=\pmatrix{
h & 0 \cr
0 & h^T \cr
}
\]
with the matrix transposition $^T$.
We include disorder by a random 
variable $M$ with mean ${\bar m}$. For our numerical transfer-matrix calculation we use a box distribution with width $W$.
For simplicity we choose the Dirac point, where $C=0$ and $D=0$. The main feature
is that there are two bands that touch each other at a spectral node $k=0$ if $M=0$, whereas $M\ne 0$ opens a gap $\Delta=2|M|$. 
Thus, a random $M$ creates a random gap. Our aim is to calculate the localization length $\Lambda$ of the eigenstate $\psi$ 
at energy $E=0$ which satisfies $h\psi=0$ and the transition probability of a moving particle. The two block Hamiltonians
$h(\bk)$, $h^*(-\bk)$ act on two separate spaces with the same localization properties. Therefore, it is sufficient to study 
just one of them.

\subsection{Localization length %: Numerical results for random gap
}
\label{sect:ll}

The localization length $\Lambda$ of the eigenstates of Hamiltonian (\ref{eq:plain-hamiltonian}) can be calculated numerically 
within a transfer-matrix approach. For this purpose the continuous Hamiltonian must be discretized in space (cf. Appendix 
\ref{app:nta}). Then the transfer-matrix $T_l$ of the eigenvalue problem $\psi_{l+1} = h^{Y} \psi_l + h^{D} \psi_{l-1}$ 
(cf. Eqs. (\ref{matrices2}), (\ref{eq:transfer-matrix-2d})) reads
\beq
T_l=\pmatrix{
h^Y & h^D \cr
1 & 0 \cr
}
\ ,
\eeq
which enables us to evaluate the Lyapunov exponents of the wavefunction \cite{Pichard1981, MacKinnon1983}.
With the initial values $\psi_0$ and $\psi_1$ the iteration of Eq.~(\ref{eq:transfer-matrix-2d}) 
provides the wavefunction $\psi_L$ at site $L$ by applying the product matrix 
\begin{equation}
 M_L = \prod_{l=1}^{L} T_l \, .
\end{equation}
For a random Hamiltonian this is a product of random matrices that satisfies Oseledec's theorem \cite{oseledec}. The latter 
states that there exists a limiting matrix 
\begin{equation}
\Gamma= \lim_{L \rightarrow \infty} (M_L^{\dagger} M_L)^{1/2L}  \, . 
\label{eq:oseledec}
\end{equation}
The eigenvalues of $\Gamma$ are usually written as a diagonal matrix with exponential functions $\exp(\gamma_i)$, 
where $\gamma_i$ is the Lyapunov exponent (LE). 
Adapting the numerical algorithm described in \cite{MacKinnon1983}, the whole Lyapunov spectrum can be calculated and 
the smallest LE is identified with the inverse localization length $1/\Lambda$ \cite{Pichard1981}. 
$\Lambda$ increases with the system width $M$ according to a power law
$
%\begin{equation}
 \Lambda \propto M^{\alpha} 
%\, ,
%\label{eq:LL-power-law}
%\end{equation}
$,
where $\alpha>1$ ($\alpha<1$) in the regime of extended (localized) states, and $\alpha=1$ in the critical regime.
For the exponentially localized regime we expect $\Lambda \propto const$. 
According to the one-parameter scaling theory by MacKinnon~\cite{MacKinnon1981},
the normalized localization length $ \tilde{\Lambda}=\Lambda/M$, being a function of disorder strength $W$ and system width $M$,
depends only on a single parameter:
\begin{equation}
 \tilde{\Lambda}(M,W) = f(\xi(W)/M) \, ,
\label{eq:scaling-sol}
\end{equation}
where $\xi$ is a characteristic length of the system generated by disorder. 
Thus, any change of disorder strength $W$ can be compensated by a change of the system width $M$.
If there is a scale-invariant point $W_c$ we can expand $\tilde{\Lambda}$ in its vicinity by assuming a
power law with critical exponent $\nu$ of the correlation length as $\xi = |W-W_c|^{-\nu}$. Then we have \cite{MacKinnon1983}
\beq
 \ln  \tilde{\Lambda} = \ln  \tilde{\Lambda}_c + \sum_{s=1}^{S} A_s \left(|W-W_c|M^{1/\nu}\right)^s \\
 =  \ln  \tilde{\Lambda}_c + \sum_{s=1}^{S} A_s \left(\frac{\xi}{M}\right)^{-s/\nu}\, .
\label{eq:fit-exponent}
\eeq
%Comparing the latter with Eq.~(\ref{eq:scaling-sol}), the scaling function $\xi$ can be interpreted as the characteristic length scale. 

\subsection{Transition probability %Dynamics
}
\label{sect:TP}

The motion of a quantum particle from site $\br'$ to site $\br$ during the time $t$
is described by the transition probability
\beq
P_{\br\br'}(t)=|\langle\br|\exp(-iHt)|\br'\rangle|^2
%P_{\br,\br'}(i\epsilon)=\frac{K_{\br,\br'}(i\epsilon)}{\sum_{\br}K_{\br,\br'}(i\epsilon)}
\ .
\eeq
If we assume that $P_{\br\br'}(t)$ describes diffusion, we can obtain the mean square displacement 
with respect to $\br'=0$ from the diffusion equation
\beq
\langle r_k^2\rangle=\sum_\br r_k^2 P_{\br,0}(t)=Dt
\ ,
\eeq
which, after applying a Laplace transformation, becomes
\beq
\sum_\br r_k^2\int_0^\infty P_{\br,0}(t)e^{-\epsilon t} dt %=\int_0^\infty Dt e^{-\epsilon t} dt
=\frac{D}{\epsilon^2}
\ .
\label{diff2}
\eeq
Using the Green's function $G_{\br\br'}(z)=(H-z)^{-1}_{\br\br'}$, we obtain for large distances $|\br-\br'|$
and $\epsilon\sim 0$
\beq
\int_0^\infty P_{\br\br'}(t)e^{-\epsilon t}dt
\sim\int_{E_0}^{E_F}\langle |G_{\br\br'}(E+i\epsilon)|^2\rangle_d dE
= \int_{E_0}^{E_F}\langle G_{\br\br'}(E+i\epsilon)G_{\br'\br}(E-i\epsilon)\rangle_d dE
\ ,
\label{gf2}
\eeq
where $\langle ...\rangle_d$ is the average with respect to disorder that is causing scattering.
$E_0$ is the lower band edge and ${\rm Tr}_4(...)$ is the trace
with respect to the $4$ spinor components. The second equation is due to the fact that the Hamiltonian 
is Hermitean. %: $H^*=H^T$.
Then we get with $\br'=0$ from Eq. (\ref{gf2}) for the diffusion coefficient at the energy $E$
\beq
%D=\epsilon^2\sum_\br r_k^2\int_0^\infty P_{\br,0}(t)e^{-\epsilon t} dt 
%=\int_0^\infty Dt e^{-\epsilon t} dt=\frac{D}{\epsilon^2}
D(E)\sim\lim_{\epsilon\to0}\epsilon^2\sum_\br  r_k^2\langle G_{\br0}(E+i\epsilon)G_{0\br}(E-i\epsilon)\rangle_d
\label{diff_c0}
\eeq
with $D=\int_{E_0}^{E_F}D(E)dE$ in Eq. (\ref{diff2}). 

%Later we will need the diffusion coefficient only at the Fermi energy. 

According to Eq. (\ref{diff2}),
diffusion requires a long range correlation %$K_{\br,\br'}(i\epsilon)$ 
for small $\epsilon$ in Eq. (\ref{gf2}). 
Anderson localization, on the other hand, is characterized by an exponentially decaying correlation. 
A natural approach to study
the latter for strong randomness would be a hopping expansion in (\ref{diff_c0}).
%of $K_{\br,\br'}(i\epsilon)$. 
Unfortunately, such
an expansion is plagued by poles on both sides of the real axis. This problem can be avoided if we focus on the most
relevant contributions of the randomly fluctuating product of Green's functions $G_{\br,\br'}(i\epsilon)G_{\br',\br}(-i\epsilon)$.
They are associated with the underlying chiral symmetry. These fluctuations have been studied previously in Ref. \cite{ziegler12}, 
where the large scale behavior was found to be associated with the Grassmann integral
\beq
%\int G(e^{{\hat S}}{\hat Q}_0e^{-{\hat S}}) J^{-1}{\cal D}[\varphi]
K_{\br \br'}=\langle G_{\br0}(E+i\epsilon)G_{0\br}(E-i\epsilon)\rangle_d\approx K_0 %4\frac{\eta^2}{g^2}
\int\varphi_\br \varphi_{\br'}' J{\cal D}[\varphi,\varphi']
\label{fint6a}
\eeq
with ${\cal D}[\varphi,\varphi']=\prod_\br d\varphi d\varphi'$ and with the Jacobian
\beq
J=\frac{1}{detg(H_0+i\epsilon+i\eta {\hat U}^2)} , \ \ \ H_0=\langle H\rangle , \ \ \ {\hat U}_\br=\pmatrix{
{\bf 1}+2\varphi_\br\varphi_\br' & -2\varphi_\br\sigma_1 \cr
-2\varphi_\br'\sigma_1 & {\bf 1}-2\varphi_\br\varphi_\br' \cr
} 
\ .
\label{jacobian2a}
\eeq
The Jacobian appears since we have restricted the integration over randomness to those degrees of freedom which
are associated with a global symmetry of the system. It is written in terms of a graded determinant $detg$, 
where the latter is expressed by conventional determinants in the relation
\[
detg\pmatrix{
A & \Theta \cr
{\bar \Theta} & B \cr
}=\frac{\det(A)}{\det(B)}\det({\bf 1}-\Theta B^{-1}{\bar \Theta}A^{-1})
\ .
\]
%The reduction to an integral over Grassmann variables reflects the fact that the long range part of the correlation 
%is associated with spontaneous symmetry breaking, where the integration is restricted to a chiral symmetry group \cite{ziegler97}.
%Then the Jacobian $J$ is generated by the projection of the integration onto this symmetry group.
The parameter $\eta$ is
the scattering rate, which can be considered as an external parameter that is either calculated in self-consistent
Born approximation \cite{ziegler09} or is taken from experimental measurements \cite{pallecchi11}. In any case, the scattering rate
increases with increasing disorder.

The relation between the correlation function $K_{\br\br'}$ %=\langle G_{\br,\br'}(i\epsilon)G_{\br',\br}(-i\epsilon)\rangle$ 
and the integral in Eq. (\ref{fint6a}) is based on two facts. Firstly, we have a large freedom to choose a distribution of
the random Green's function with the same expectation value. Secondly, by choosing a proper distribution we find a saddle-point
approximation for the corresponding integration. This procedure was described in detail in Refs. \cite{ziegler12,ziegler09}, leading eventually 
to Eq. (\ref{fint6a}). As a result we have been able to avoid the spurious singularities, which appear when we apply a hopping expansion and 
integrate with respect to the random term of the Hamiltonian.

The expression in Eq. (\ref{jacobian2a}) enables us to rewrite $J$ for weak scattering ($\eta\ll 1$) as
\beq
J=1/detg\left[{\bf 1}+i\eta ({\hat H}_0+i\epsilon)^{-1}{\hat U}^2\right]
\equiv 1/detg({\bf 1}+i\eta {\hat G}_0{\hat U}^2)
\label{wsc}
\eeq
and for strong scattering ($\eta\gg 1$) as
\beq
J=1/detg\left[{\bf 1}+\frac{1}{i\eta}({\hat H}_0+i\epsilon){\hat U}^{-2}\right]
\label{ssc}
\eeq
since $detg({\hat U}^2)=1$. These expressions can be used to employ an expansion in powers of $\eta$ or $1/\eta$, 
respectively.
The expression in Eq. (\ref{wsc}) has been treated previously. It leads to diffusion, 
where the correlation function is a diffusion propagator \cite{ziegler97}. In Sect. \ref{sect:strong} 
we will extend the previous work to the regime of strong scattering, employing an expansion in powers of
$1/\eta$ for the expression (\ref{ssc}).

\section{Numerical Results: scaling of the localization length}

\begin{figure}[ht]
\includegraphics[scale=0.62]{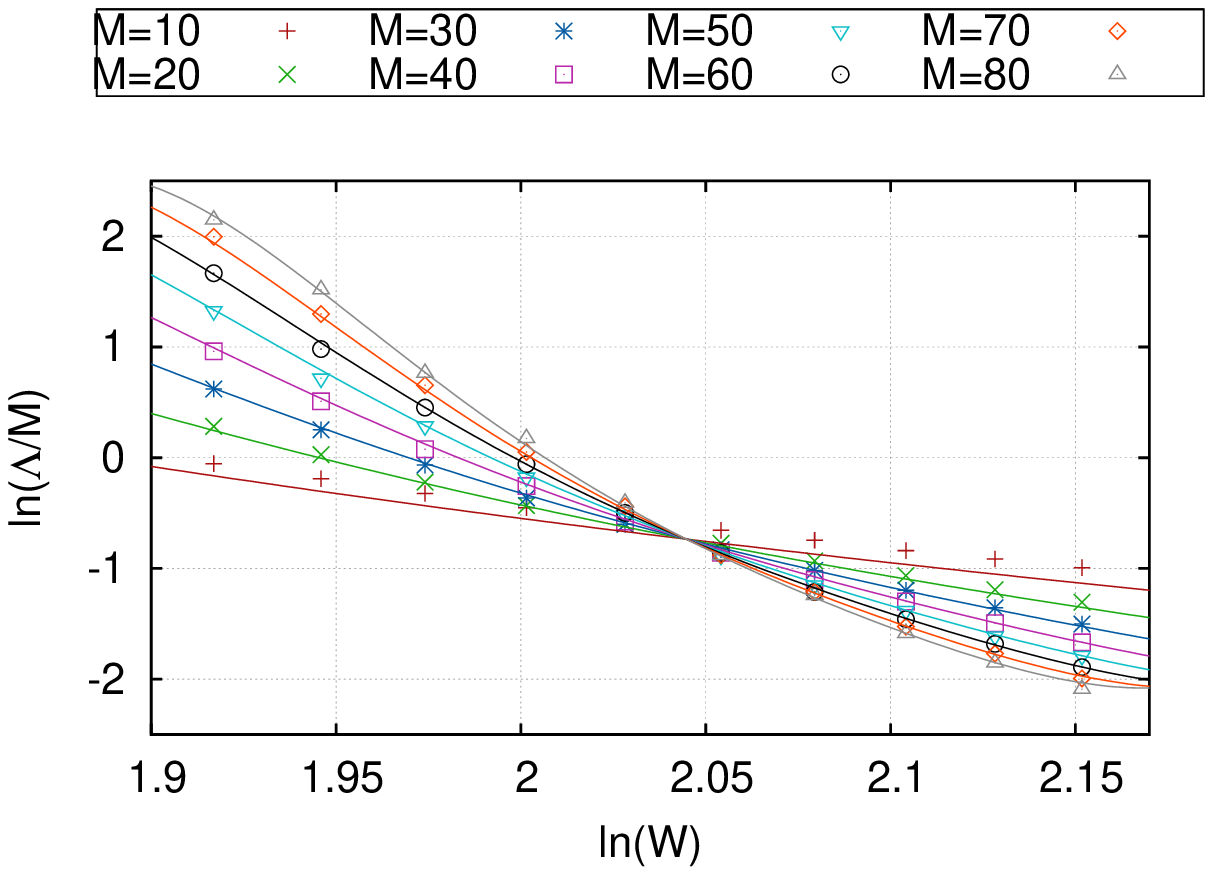} 
\includegraphics[scale=0.62]{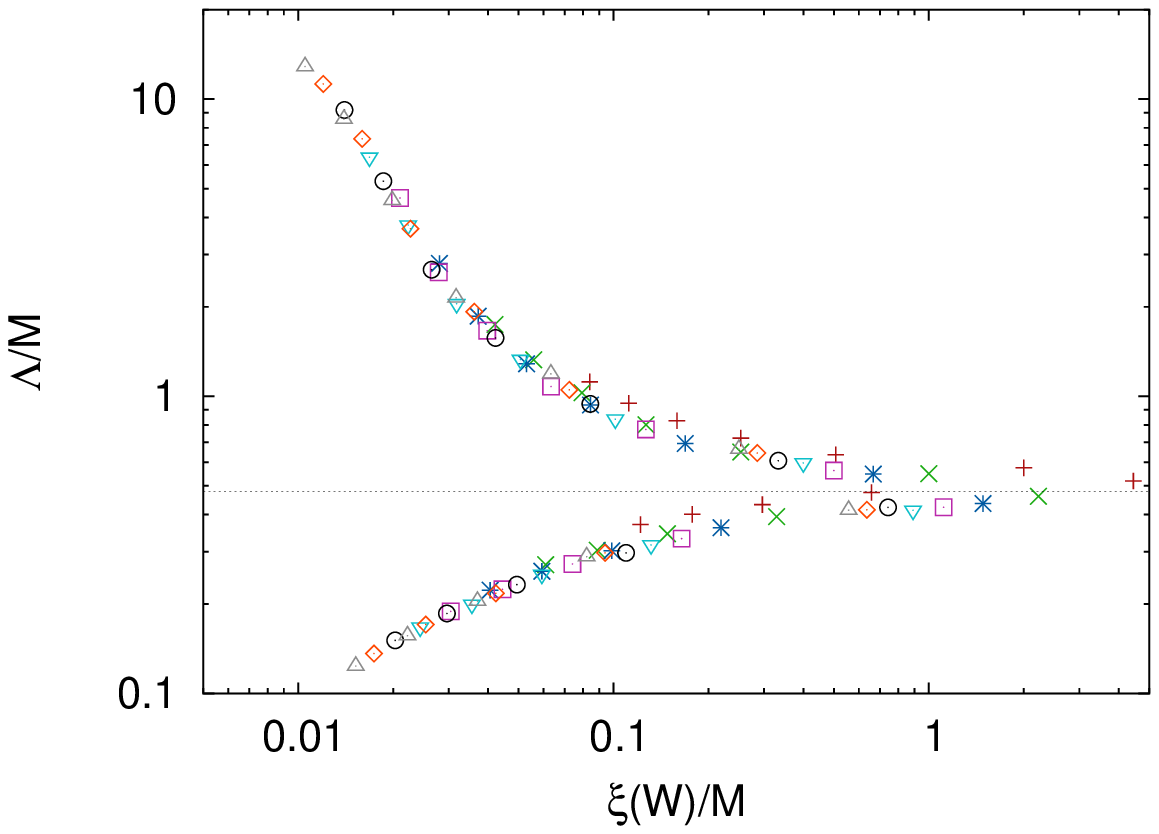}
\caption{Numerical evidence for a localization transition in two dimensions. The scaling behavior of the normalized localization
length ${\bar\Lambda}$ as a function of increasing disorder $W$ is plotted here for $\bar{m}= 0.8$ and $\delta=0.5$.
Left panel: Fit to Eq.~(\ref{eq:fit-exponent}) near the critical point.
Right panel: Rescaled normalized localization length ${\bar\Lambda}$ near the critical point. 
%The plot contains more data points than used for the fitting procedure to show the validity of one parameter scaling.
}
\label{fig:rescaled-nll-gap0-delta05_a}
\end{figure}

Now we return to the method described in Sect. \ref{sect:ll} and calculate the localization length $\Lambda$.
Our calculation for strong randomness (i.e. large $W$) provides a critical value $W_c$, where the system is 
delocalized (localized) for $W<W_c$ ($W>W_c$). Around the critical value $W_c$ we observe one-parameter 
scaling behavior for the normalized localization length ${\bar \Lambda}$, as described in Sect. \ref{sect:ll}.
Some results are depicted in Fig. \ref{fig:rescaled-nll-gap0-delta05_a} and the results of the fitting
procedure are listed in Table \ref{table:critical-values-gap02-delta05}.
This behavior is indicative of an Anderson transition.

\begin{table}[ht]
\centering
    \begin{tabular}{ l  c  c  c}
    average gap $\bar{m}$ & $0$ & $0.2$ & $0.8$ \\ \hline\hline
	exponent $\nu$ & $1.299\pm 0.066$ & $1.397 \pm 0.069$ & $1.451 \pm 0.024$ \\ 
	critical disorder $W_c$ & $7.668\pm0.008$ & $7.629 \pm 0.015$ & $7.727 \pm 0.01$ \\
	%$\Lambda_c$ & $0.447\pm0.005$ & $0.517 \pm 0.009$   & $0.479 \pm 0.007$  \\
	disorder range & $7.35\leq W \leq 7.8$ & $7.1\leq W \leq 8.0$  & $6.6\leq W \leq 8.4$ \\ 
	system sizes & $30 \leq M \leq 80$ & $20 \leq M \leq 80$ & $20 \leq M \leq 80$  \\
	\hline
    \end{tabular}
    \caption{Critical values for $\delta=0.5$ obtained from fitting the data to Eq.~(\ref{eq:fit-exponent}).}
    \label{table:critical-values-gap02-delta05}
\end{table}

\section{Analytic Results: strong scattering expansion}
\label{sect:strong}

\begin{figure}[ht]
\begin{center}
\includegraphics[scale=0.3]{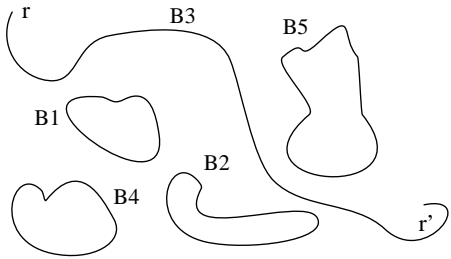} 
\caption{Typical contribution to the $1/\eta$ expansion of $Z$. %the Jacobian.
There are five connected clusters B1,...,B5 which are disconnected of each other.
In particular, there is a random walk B3 from site $\br$ to site $\br'$, the only
contribution to the correlation function $K_{\br \br'}$ in Eq. (\ref{exp3}).
}
\label{fig:exp}
\end{center}
\end{figure}

Eq. (\ref{fint6a}) is a convenient starting point to study transport properties with the Jacobian
\beq
J=%{\bar J}
\exp\left\{-{\rm Trg}\left[\log\left({\bf 1}+\frac{1}{i\eta}({\hat H}_0+i\epsilon){\hat U}^{-2}
\right)\right]
\right\}
%\sim {\bar J}\exp[-{\rm Tr}(\gamma_+\varphi \varphi'-\varphi \gamma_-\varphi'+\gamma_+\varphi \gamma_-\varphi')]
\ ,
\label{jacobian5}
\eeq
where the graded trace $Trg$ is with respect to the four-dimensional spinor space and the position $\br$.
It is related to the conventional trace by
\[
Trg\pmatrix{
A & \Theta \cr
{\bar \Theta} & B \cr
}=Tr A- Tr B
\ .
\]
The integral representation of the correlation function $K_{\br \br'}$ in Eq. (\ref{fint6a}) with the Jacobian in 
Eq. (\ref{jacobian5}) enables us to study the regime of strong scattering (i.e. $\eta\gg 1$) by applying a $1/\eta$
expansion. 
This allows us to rewrite the correlation function %in Eq. (\ref{fint6a}) 
as
\[
K_{\br \br'}\approx K_0 %4\frac{\eta^2}{g^2}
\int\varphi_\br \varphi_{\br'}' 
\exp\left\{-{\rm Trg}\left[\log\left({\bf 1}+\frac{1}{i\eta}({\hat H}_0+i\epsilon){\hat U}^{-2}\right)\right]\right\}
{\cal D}[\varphi,\varphi']
\]
\beq
=K_0 %4\frac{\eta^2}{g^2}
\frac{\partial}{\partial\alpha}\int 
\exp\left\{\alpha\varphi_\br \varphi_{\br'}' 
-{\rm Trg}\left[\log\left({\bf 1}+\frac{1}{i\eta}({\hat H}_0+i\epsilon){\hat U}^{-2}\right)\right]\right\}
{\cal D}[\varphi,\varphi']\big|_{\alpha=0}
\label{fint6b}
\eeq
and to expand the exponential function as
\[
=K_0 %4\frac{\eta^2}{g^2}
\frac{\partial}{\partial\alpha}\sum_{l\ge0}\frac{1}{l!}\int
\left(\alpha\varphi_\br \varphi_{\br'}'
-{\rm Trg}\left[\log\left({\bf 1}+\frac{1}{i\eta}({\hat H}_0+i\epsilon){\hat U}^{-2}\right)\right]
\right)^l
{\cal D}[\varphi,\varphi']\big|_{\alpha=0}
\]
\beq
= K_0 %4\frac{\eta^2}{g^2}
\frac{\partial Z}{\partial\alpha}\Big|_{\alpha=0}
\ .
\eeq
Here we have used the expression
\[
Z=\sum_{l\ge0}\frac{1}{l!}\langle (\sum_jA_j)^l\rangle
\ ,
\]
where $\sum_jA_j$ is the expansion of 
$\alpha\varphi_\br \varphi_{\br'}'-{\rm Trg}\left[\log\left({\bf 1}+\frac{1}{i\eta}({\hat H}_0+i\epsilon){\hat U}^{-2}\right)\right]$:
\[
\sum_j A_j= %\alpha\varphi_\br \varphi_{\br'}' -{\rm Tr}\log[{\bf 1}+h\varphi \varphi'+\varphi h\varphi'-h\varphi h\varphi']
\alpha\varphi_\br \varphi_{\br'}' +\sum_{j\ge 1}\frac{(-1)^j}{j(i\eta)^j}{\rm Trg}
\left([({\hat H}_0+i\epsilon){\hat U}^{-2}]^j\right)
\]
and the average is with respect to the normalized integral:
\[
\langle ...\rangle=\frac{1}{\cal N}\int ... {\cal D}[\varphi,\varphi']
\ .
\]
Using the fact that the factors of the product $A_{j_1}A_{j_2}\cdots A_{j_l}$ can be reorganized as products of 
connected clusters $\{ B_k\}$ (cf. Appendix \ref{app:lct}), we obtain from the Linked Cluster Theorem
\beq
\frac{\partial Z}{\partial\alpha}=Z\frac{\partial \log Z}{\partial\alpha}
=Z\frac{\partial }{\partial\alpha}\sum_k \langle B_k\rangle
\ .
\label{exp3}
\eeq
Thus only those expressions $\langle B_k\rangle$ contribute that contain $\alpha$.
These contributions form random walks from site $\br$ to site $\br'$ with the discrete hopping term of Eqs. 
(\ref{discrete}) and (\ref{disc-dirac}) (cf. Fig. \ref{fig:exp}). They can be estimated as 
\beq
\Big|\frac{\partial }{\partial\alpha}\sum_k \langle B_k\rangle\Big|_{\alpha=0}\Big|
\le \sum_{l\ge |\br -\br'|}\frac{1}{\eta^l} |Tr_4[({\hat H}_0^l)_{\br,\br'}]|
\le const. (4/\eta)^{|\br -\br'|}
\ ,
\eeq
where the factor 4 is due to the two dimensional random walk. % and $\mu=\max(v_F,\delta)$. 
Thus, we need $\eta>4$ (in units of $\hbar v_F/a$ with lattice constant $a$) in order to have
an absolutely convergent series and an exponential decay of the correlations. The latter 
describes Anderson localization, according to our discussion in Sect. \ref{sect:TP}. 

\section{Discussion and conclusion}
%-- essential points (requires some additional work and discussion):
%1)
Our analytic calculation supports the numerical result of a localized phase at sufficiently strong disorder.
Here it should be noticed that the calculations are based on different quantities, namely the localization
length $\Lambda$ and the exponential decay of the average transition matrix $K_{\br \br'}$. Since the
localization length is self-averaging according to Eq. (\ref{eq:oseledec}), it is expected that this
quantity should be very robust in a real system. On the other hand, the localization length is difficult
to measure directly in an experiment. Therefore, the transition matrix is more accessible because it is
related to the conductivity by the Einstein relation
\[
\sigma_{xx}\approx \epsilon^2\frac{e^2}{h}\sum_\br x^2 K_{\br 0}
\ ,
\] 
where $x$ is the direction of the position $\br$, in which the external electric field is applied.
In the DC limit $\epsilon\to 0$ the conductivity vanishes when $K_{\br 0}$ decays exponentially. This
is in stark contrast to the weak scattering case where the expansion in powers of $\eta$ gives a 
diffusion propagator \cite{ziegler09}
\[
{\tilde K}_\bq\propto \frac{1}{\epsilon +Dq^2}
\]
with the diffusion coefficient $D$.
After Fourier transformation $\bq\to\br$ this expression gives a correlation function that decays like 
$\sim r^{-1/2}$. Moreover, it gives a finite non-vanishing DC conductivity, since 
\[
\sum_\br x^2 K_{\br 0} =-\frac{\partial {\tilde K}_\bq}{\partial q_x^2}\Big|_{\bq=0}\propto\frac{2D}{\epsilon^2}
\ .
\]
On the surface of a typical topological insulator we expect substantial scattering due to
disorder \cite{zhang11}. Our results indicate that the suppressed backscattering may not be able 
to prevent the localization
of surface states. Therefore, it might be crucial for the appearance of a metallic behavior to reduce the disorder 
on the surface. In this case it could even be possible to observe an Anderson transition from extended to localized
states, as our results indicate. Our calculation gives a rough estimate for the localized behavior in which
the scattering rate must be larger than the bandwidth of the system without disorder. A similar transition was 
also observed in a numerical study of the conductivity in disordered graphene by Zhang et al. \cite{zhang09}.
However, we cannot confirm their interpretation as a Kosterlitz-Thouless transition because we find a
power law for the localization length.  

In conclusion, we have studied a model for surface states on a topological insulator. Contrary to the assumption
that suppressed backscattering may always create a metallic phase, we have found that the surface states are 
localized for strong scattering by disorder. For weak scattering, however, there is a metallic behavior and a 
phase transition from a delocalized to a localized phase when the disorder strength is increased. The transition 
is characterized by one-parameter scaling of the normalized localization length with a non-universal exponent.

\appendix

\section{Numerical transfer-matrix calculation}
\label{app:nta}

A numerical treatment of the Dirac Hamiltonian requires a discretization in space.
However, the naive discretization through replacing the differential operator by a 
difference operator leads to additional new nodes, which is often called fermion doubling 
or multiplication ~\cite{Susskind1977}. In real space there are two methods to circumvent 
this problem~\cite{Stacey1982,beenakker08,beenakker10}. One that we will adopt in this
section goes back to an idea of Susskind.
%The Dirac equation for free particles reads in matrix notation
%\begin{equation}
% -\alpha \begin{pmatrix}0 & i\partial_x + \partial_y \\ i\partial_x - \partial_y  & 0     \end{pmatrix}
%\begin{pmatrix}\varphi_1 \\ \varphi_2\end{pmatrix}= E \begin{pmatrix}\varphi_1 \\ \varphi_2\end{pmatrix} \, .
%\end{equation}
We start with discretizing the differential operator in an anti-symmetric way
\begin{equation}
\partial_x f(x)\approx \frac{1}{2\Delta} (f_{l+\Delta} - f_{l-\Delta} )
 \ ,
\label{discrete}
\end{equation}
where $\Delta$ is the lattice constant which we set to one %$\hbar v_F$ 
in the following. The discrete Dirac equation for $m=0$ and with $\hbar v_F=1$ a 
then takes the form
\beq
 -\frac{i}{2} \sigma_1 \left\lbrace \psi_{l+1,n} - \psi_{l-1,n}  
\right\rbrace -\frac{i}{2} \sigma_2 \left\lbrace\psi_{l,n+1} - \psi_{l,n-1}  \right\rbrace = E\sigma_0\psi_{l,n} \, 
\label{disc-dirac}
\eeq
with lattice points given by the integer coordinates $(l,n)$. Fourier transformation leads to eigenvalues 
$E=\pm \sqrt{sin(k_x)^2 + sin(k_y)^2}$ which have four Dirac cones in the Brillouin zone corresponding to four Dirac fermions. 
In order to open a gap at three of them we introduce a lattice operator which acts on a wave function as~\cite{Ziegler1996}
\begin{equation}
 \hat{B} \, \psi_{l,n} = \frac{1}{2}\left\lbrace \psi_{l+1,n} + \psi_{l-1,n} + \psi_{l,n+1} + \psi_{l,n-1}  \right\rbrace \, .
\end{equation}
The discretized form of the Hamiltonian~(\ref{eq:plain-hamiltonian}) for uniform gap now reads
\begin{equation}
  h= \sin(k_x)\sigma_1 - \sin(k_y)\sigma_2 + \left[m+\delta (cos(k_x)+cos(k_y)-2)\right] \sigma_3
% \begin{pmatrix}
%      m+\delta (cos(k_x)+cos(k_y)-2)  & \sin(k_x) + i\sin(k_y) \\  \sin(k_x) - i\sin(k_y) & -m-\delta (cos(k_x)+cos(k_y)-2)
%     \end{pmatrix}
    \label{main_ham}
\ ,
\end{equation}
which gives $h(\bk)$ of Eq. (\ref{eq:plain-hamiltonian}) in the continuum limit and has the dispersion
\begin{equation}
 E = \pm \sqrt{sin(k_x)^2 + sin(k_y)^2 + (m+\delta cos(k_x)+\delta cos(k_y)-2\delta )^2} \, .
\label{eq:dispersion}                                                                                            
\end{equation}
For $m=0, \delta\ne0$ there is a node at $k_x=k_y=0$ and three additional nodes for $m=0, \delta=0$ at $k_x,k_y=\pm\pi$ %(cf. Fig.~\ref{fig:dispersion-vergleich}).
Using this model node degeneracy can be lifted via the parameter $\delta$.

We absorb the index $n$ with the help of matrix representation and write for the wave function
\begin{equation}
  \psi_{l+1} = h^{Y} \ \psi_l + h^{D} \ \psi_{l-1} \, .
\end{equation}
Each spinor component is now a $M$-component vector, where $M$ is the width of a strip and thus $n=1,2,...,M$. The matrices $h^{Y}$, $h^{D}$ read
\[
 h^{Y}_{n,n} = 2 S^{-1}\left[  E\, \sigma_0  + (2\delta-m)\sigma_3 \right] \ \ \  h^{Y}_{n,n+1} = S^{-1} \left[i\sigma_2 -\delta \sigma_3  \right]
\]
 \beq
 h^{Y}_{n,n-1} = -  S^{-1}\left[ i\sigma_2 + \delta\sigma_3  \right]  \ \ \  h^D_{n,n} = - S^{-1}\left[i \sigma_1  + \delta \sigma_3  \right] \, 
\label{matrices2}
\eeq
with $S=-i\sigma_1 + \delta \sigma_3$ and where $h^{Y}$ has periodic boundary conditions in the $y$-direction.
This matrix structure allows us to construct a transfer matrix $T_l$ through the equation \cite{MacKinnon1983}
\begin{equation}
\pmatrix{
\psi_{l+1} \cr
\psi_{l} \cr
} = \pmatrix{
h^Y & h^D \cr
1 & 0 \cr
}\pmatrix{
\psi_l \cr
\psi_{l-1}\cr
} \equiv T_l \pmatrix{
\psi_l \cr
\psi_{l-1} \cr
}\, .
\label{eq:transfer-matrix-2d}
\end{equation}

\section{Linked Cluster Theorem}
\label{app:lct}

We must organize the $1/\eta$ expansion in order to extract the spatial decay of the correlation function $K_{\br \br'}$.
For this purpose we employ the Linked Cluster Theorem \cite{glimm}. The latter can be formulated for the expression
\beq
\frac{1}{l!}\langle (\sum_jA_j)^l\rangle
=\frac{1}{l!}\sum_{j_1,j_2,...,j_l}\langle A_{j_1}A_{j_2}\cdots A_{j_l}\rangle
\ .
\label{summ2}
\eeq
The product of the $A_iA_j$ is called disconnected (unlinked) if the two factors do not share any Grassmann variable.
This would lead to $\langle A_iA_j\rangle=\langle A_i\rangle\langle A_j\rangle$. Otherwise they are called connected
(linked) and we would have $\langle A_iA_j\rangle\ne\langle A_i\rangle\langle A_j\rangle$. In the sum (\ref{summ2})
we combine for a given set $j_1,j_2,...,j_l$ all connected factors in products $\{B_k\}$ such that 
\beq
\langle A_{j_1}A_{j_2}\cdots A_{j_l}\rangle =\langle B_{k_1}\rangle\langle B_{k_2}\rangle\cdots \langle B_{k_n}\rangle \ \ \ (n\le l)
\ ,
\label{factors}
\eeq
where the new indices $k_1,...,k_n$ refer to the indices $j_1,...,j_l$ of the combined factors $A_j$. Now we must reorganize the summation.
A permutation of the $j_1,j_2,...,j_l$ gives the same expression for (\ref{factors}). Therefore, the summation with respect to the permutations
contributes only a factor $l!$. On the other hand, we allow also a permutation of the $k_1,k_2,...,k_n$, which would also
leave the expression (\ref{factors}) invariant. Consequently, we must divide the summation with respect to these $n$ permutations
by $n!$. This gives us eventually
\[
Z=\sum_{l\ge0}\frac{1}{l!}\langle (\sum_j A_j)^l\rangle
=\sum_{n\ge0}\frac{1}{n!}(\sum_k \langle B_k\rangle)^n
=\exp\left(\sum_k \langle B_k\rangle\right)
\ ,
\]
which is the Linked Cluster Theorem, since the $B_k$ are connected according to our construction.

\end{document}